\documentclass[12pt]{article}
\usepackage[english]{babel}
\usepackage[normalem]{ulem}
\usepackage{multirow}
\usepackage{booktabs}
\usepackage[text={15.5cm,22.5cm}]{geometry}
\usepackage{graphicx}
\usepackage[dvipsnames]{xcolor}
\usepackage{amsmath}
\usepackage{amssymb}
\usepackage{slashed}
\usepackage{mathrsfs}
\usepackage{yfonts}
\usepackage{cite}

\newcommand{\hatP}{{\widehat P}}
\newcommand{\hatQ}{{\widehat Q}}

\begin{document}
	
	\title{\vspace{-0.8cm}
		{\normalsize
			\flushright TUM-HEP 1267/20\\}
		\vspace{1cm}
		\bf Two-loop renormalization group equations for right-handed neutrino masses and phenomenological implications
		\\ [8mm]}
	
	\author{Alejandro Ibarra$^1$, Patrick Strobl$^1$, Takashi Toma$^{2,3}$\\[2mm]
		{\normalsize\it $^1$ Physik-Department, Technische Universit\"at M\"unchen, }\\[-0.05cm]
		{\normalsize\it James-Franck-Stra\ss{}e, 85748 Garching, Germany}\\[2mm]
		{\normalsize\it $^2$ Institute of Liberal Arts and Science, Kanazawa University, }\\[-0.05cm]
		{\normalsize\it Kakuma-machi, Kanazawa 920-1192, Japan}\\[2mm]
		{\normalsize\it $^3$Department of Physics, McGill University,}\\
		{\normalsize\it 3600 Rue University, Montr\'{e}al, Qu\'{e}bec H3A 2T8, Canada}
	}

	\date{}
	\maketitle
	\thispagestyle{empty}
	\vskip 1.5cm
	\begin{abstract}
		We calculate the two-loop beta functions of the right-handed neutrino mass matrix in the Standard Model extended with right-handed neutrinos. We show that two-loop quantum effects induced by the heavier right-handed neutrinos can induce sizable contributions (sometimes dominant) to the physical masses of the lighter right-handed neutrinos. These effects can significantly affect the masses of the active neutrinos in the seesaw mechanism and the low energy phenomenology. 
	\end{abstract}
	
	\newpage
	\section{Introduction}
	\label{sec:intro}

	Neutrino oscillation experiments have demonstrated the existence of at least two non-vanishing neutrino masses. The measured mass differences are much smaller than the electroweak symmetry breaking scale or the mass of any other Standard Model fermion. Besides, the misalignment between the neutrino interaction and mass eigenstates is qualitatively different to the one measured for the quarks, and the mass hierarchy between the largest and next-to-largest active neutrino masses is milder than the one observed in the quark or charged lepton sectors. All these facts suggest that the mechanism generating neutrino masses could be fundamentally different to the one generating quark or charged lepton masses. 
	
	Neutrinos are the only electrically neutral fermions in the Standard Model, and therefore the only Standard Model particle that admits a Majorana mass. It is then conceivable that the differences between the neutrino and the quark or charged lepton parameters could be related to the Majorana nature of the neutrinos.
	
	In an effective theory approach, Majorana masses for the active neutrinos can be incorporated into the Standard Model by adding a $SU(2)_L\times U(1)_Y$ gauge invariant dimension-5 operator, the so-called Weinberg operator~\cite{Weinberg:1979sa}. This description is only valid up to a cut-off scale, where the effective theory must be replaced by a renormalizable theory involving new degrees of freedom. Interestingly, current experiments indicate that this effective description of neutrino masses in terms of a Weinberg operator cannot be valid up to the Planck scale, $M_{\rm P}=1.2\times 10^{19}$ GeV. Should this be the case, all neutrino masses should be $m_\nu \lesssim \langle H^0\rangle ^2/M_{\rm P}\sim 10^{-6}$ eV~\cite{Barbieri:1979hc}, which is too small to explain the measured mass splittings (here, $ \langle H^0\rangle=174$ GeV is the Higgs vacuum expectation value). Therefore, current experiments demonstrate that either the fundamental energy cut-off of Nature is smaller than the Planck scale, or that new particles exist with mass between the electroweak and the Planck scale, which are responsible for the generation of neutrino masses.
	
	One of the simplest ultraviolet completions to the Weinberg operator consists in adding to the Standard Model particle content several right-handed neutrinos, with mass much larger than the electroweak symmetry breaking scale, but smaller than the Planck scale~\cite{Minkowski:1977sc, Yanagida:1979as, GellMann:1980vs}. In this framework, the heavy neutrinos are integrated-out at the energy scales relevant to oscillation experiments, generating a Weinberg operator which is suppressed by the mass scale of the heavy neutrinos. Correspondingly, the active neutrino masses which arise after the electroweak symmetry breaking can be larger than $10^{-6}$ eV, thus enabling to reproduce the oscillation data by adjusting the parameters of the model. 
	
	At energy scales between the fundamental cut-off of Nature, possibly the Planck scale, and a given right-handed neutrino mass, the corresponding right-handed neutrino is a dynamical degree of freedom and can affect the values of the model parameters through quantum effects. The renormalization group equations (RGEs) of the Yukawa couplings were calculated in Ref.\cite{Cheng:1973nv} at one loop and in Ref.\cite{Machacek:1983fi} at two loops; the RGEs of the right-handed mass matrix were calculated in Ref.\cite{Casas:1999tp} at one loop. In this work we will  present for the first time the full two-loop RGE of the right-handed mass matrix in the Standard Model extended with right-handed neutrinos.
	
	We are motivated by the fact that two-loop quantum effects on the right-handed neutrino mass matrix can have a significant impact on the values of the physical parameters, in contrast to the two-loop quantum effects on the neutrino Yukawa couplings which just give small corrections to the eigenvalues and diagonalization matrices. The crucial difference between the RGE of a Majorana mass matrix and of a Yukawa matrix lies in the violation in the former case of the total lepton number. Since a given right-handed Majorana mass is (in general) not protected by any symmetry, it can receive contributions proportional to other right-handed neutrino masses. In particular, the physical mass of the lighter right-handed neutrinos can be much larger than their tree-level values due to two-loop contributions from the heavier right-handed neutrinos~\cite{Ibarra:2018dib,Aparici:2012vx}.\footnote{An analogous mechanism leads to a lower bound on the lightest active neutrino mass from two-loop quantum effects induced by the heaviest active neutrino \cite{Petcov:1984nz,Davidson:2006tg}.} In particular, it was argued in Ref.~\cite{Ibarra:2018dib} that the breaking of the lepton number at the Planck scale could naturally lead to active neutrino masses of ${\cal O}(0.1)$ eV, if one of the right-handed neutrino masses is dominated by two-loop quantum effects. 
	
	The paper is organized as follows. In Section~\ref{sec:2}, we present the two-loop RGEs for the right-handed neutrino mass matrix in the Standard Model extended with right-handed neutrinos. In Section~\ref{sec:3}, we study the implications of the two-loop RGEs for the active neutrino masses in a scenario with two right-handed neutrinos, focusing on the case where the cut-off value of the heaviest right-handed mass is close to the Planck scale and of the lightest is very small, such that its physical mass is dominated by quantum effects. In Section~\ref{sec:4} we extend the analysis to the three right-handed neutrino scenario, where again the heaviest right-handed neutrino mass is close to the Planck scale, and either one or two of the lighter right-handed neutrino masses are dominated by two-loop quantum effects. Finally, in Section~\ref{sec:5} we present our conclusions.
		
	\section{Two-loop RGEs of right-handed neutrino parameters}
	\label{sec:2}
	
	We consider an extension of the Standard Model by $n_g$ right-handed neutrinos, $N_i$, $i=1,..., n_g$. The most general renormalizable Lagrangian involving the right-handed neutrinos reads:
	\begin{align}
	{\cal L}_{N}=\frac{1}{2}\overline{N_i}i\slashed{\partial}N_i
	- Y_{\alpha i} \overline{L_\alpha} \widetilde H N_i-\frac{1}{2}M_{ij} \overline{N_i^c} N_j+{\rm h.c.}\;,
	\label{eq:Lagrangian}
	\end{align}
	where $L_\alpha~(\alpha=e,\mu,\tau)$ are the lepton doublets and 
	$\widetilde H= i \tau_2 H^*$ is the charge conjugate of the Standard Model Higgs doublet $H$.
	
	The Yukawa matrix $Y$ and the mass matrix $M$ are generated by some yet unknown mechanism at a high energy scale $\Lambda$, below which the model is well described by the Lagrangian Eq.~(\ref{eq:Lagrangian}). The parameters of the model, on the other hand, are subject to quantum effects. In this paper we focus on the effects proportional to $\log(\Lambda/M_i)$ (with $M_i$ the physical masses of the right-handed neutrinos) which can be calculated using the renormalization group equations (RGEs).
	
	The RGE of the right-handed Majorana mass matrix up to two-loops is given by
	\begin{align}
	\frac{d M}{d \log\mu} = \beta^{(1)}_M+\beta^{(2)}_M,
	\end{align}
	with $\beta_{M}^{(1)}$ and $\beta_{M}^{(2)}$ the one- and two-loop beta functions.
	The one-loop $\beta$ function was calculated in Ref.~\cite{Casas:1999tp}, while the two-loop beta function is to the best of our knowledge a new result:
	\begin{align}
	16 \pi^2 \beta^{(1)}_M=&~M  (Y^\dagger Y) + (Y^\dagger Y)^T  M, \nonumber \\
	(16\pi^2)^2\beta^{(2)}_M=&~4 (Y^\dagger Y)^T  M  (Y^\dagger Y) -  \frac{1}{4} \left[ M  (Y^\dagger Y) (Y^\dagger Y) + (Y^\dagger Y)^T (Y^\dagger Y)^T M \right]\nonumber \\
	&  +\bigg\{\frac{17}{8} g_1^2 +\frac{51}{8} g_2^2- \frac{9}{2} y_t^2-\frac{3}{2} \textrm{Tr}\left(Y^\dagger Y\right) \bigg\} \left[M  (Y^\dagger Y) + (Y^\dagger Y)^T  M\right].
	\label{eq:beta-2loop}
	\end{align}
	(We choose to work in the MS scheme.)
	Here,  $g_1$ and $g_2$ are the gauge couplings corresponding to the $U(1)_Y$ and $SU(2)_L$ symmetries and $y_t$ is the top Yukawa coupling. We have neglected the effects in the running of the remaining Standard Model Yukawa couplings. 
	
	Besides, the RGE of the neutrino Yukawa coupling is
	\begin{align}
	\frac{d Y}{d \log\mu}=\beta_{Y}^{(1)}+\beta_{Y}^{(2)},
	\end{align}
	with  $\beta_{Y}^{(1)}$ and $\beta_{Y}^{(2)}$ the one- and two-loop beta functions. These were calculated in Ref.~\cite{Machacek:1983fi} (see also \cite{Pirogov:1998tj}), and are reproduced here for completeness:
	\begin{align}
	16\pi^2 \beta_{Y}^{(1)} =&~\bigg[3 y_t^2 + \textrm{Tr}(Y^\dagger Y) - \frac{3}{4} g_1^2 - \frac{9}{4} g_2^2\bigg] Y + \frac{3}{2} Y Y^{\dagger} Y,\nonumber \\
	(16\pi^2)^2\beta_{Y}^{(2)}=&~\bigg[\frac{93}{16} g_1^2 + \frac{135}{16} g_2^2 - \frac{27}{4} y_t^2 - \frac{9}{4}\textrm{Tr}(Y^\dagger Y) - 12\lambda\bigg] Y Y^\dagger Y\nonumber\\
	&+ \bigg[6 \lambda^2 -\frac{27}{4} y_t^4 - \frac{9}{4} \, \textrm{Tr}\big((Y^\dagger Y)^2\big) - \frac{9}{4} g_1^2 g_2^2 + \frac{35}{24} g_1^4- \frac{23}{4} g_2^4\nonumber\\
	&+ \bigg(\frac{85}{24} g_1^2 + \frac{45}{8} g_2^2 + 20 g_s^2\bigg) y_t^2 + \bigg(\frac{5}{8} g_1^2 + \frac{15}{8} g_2^2\bigg) \, \textrm{Tr}(Y^\dagger Y)\bigg] Y + \frac{3}{2} Y (Y^\dagger Y)^2,
	\end{align}
	with $\lambda$ the Higgs quartic coupling and $g_s$ the $SU(3)_c$ gauge coupling.
	
	To discuss the main qualitative new features of the two-loop quantum effects on the right-handed neutrino mass matrix it is convenient to introduce the quantities
	\begin{align}
	P&=\frac{1}{16\pi^2} Y^\dagger Y\;,\nonumber \\
	{\cal G} &=\frac{1}{16\pi^2}\Big(\frac{17}{8} g_1^2+\frac{51}{8} g_2^2-\frac{9}{2} y_t^2 \Big)-\frac{3}{2} \textrm{Tr}P
	\;,\nonumber\\
	Q&=(1+{\cal G})P-\frac{1}{4}P^2.
	\label{eq:Q}
	\end{align}
	Then the RGE for the mass matrix can be recast as
	\begin{align}
	\frac{dM}{d \log\mu} = MQ + Q^TM +4 P^TMP.
	\label{eq:RGE-mass-matrix}
	\end{align}
	
	The RGEs for the right-handed neutrino mass eigenvalues and eigenvectors can be obtained following the lines of Ref.~\cite{Casas:1999tg}. We decompose the mass matrix in terms of a real diagonal matrix containing the mass eigenvalues $D_M$ and a complex unitary matrix $U_M$, such that $M=U_M^* D_M U_M^\dagger$. Then, the real and imaginary parts of the diagonal elements of the RGE are:
	\begin{align}
	\frac{dM_i}{d\log\mu}&=  2M_i {\widehat Q}_{ii}+4 \sum_k  M_k {\rm Re}\,\left({\widehat{P}}^2_{ki}\right), \label{eq:Re_diag}\\
	-2 M_i {\rm Im}\left(U_M^\dagger\frac{dU_M}{d\log\mu}\right)_{ii}&=4\sum_k M_k {\rm Im}\left(\widehat P^2_{ki}\right), \label{eq:Im_diag}
	\end{align}
	and of the off-diagonal elements are
	\begin{align}
	(M_j-M_i){\rm Re}\left(U_M^\dagger\frac{dU_M}{d\log\mu}\right)_{ij}
	&=(M_i+M_j){\rm Re}\left(\widehat  Q_{ij} \right)+4 \sum_k M_k {\rm Re}\left( \widehat P_{ki}\widehat P_{kj}\right),\label{eq:Re_off_diag}\\
	-(M_j+M_i){\rm Im}\left(U_M^\dagger\frac{dU_M}{d\log\mu}\right)_{ij}
	&=(M_i-M_j){\rm Im}\left(\widehat Q_{ij}\right)+4\sum_k M_k {\rm Im}\left(\widehat P_{ki}\widehat P_{kj}\right),\label{eq:Im_off_diag}
	\end{align}
	where $\widehat P =U_M^\dagger P U_M$ and $\widehat Q =U_M^\dagger Q U_M$.
	Notice that the hermiticity of the matrices $\widehat
	P$ and $\widehat Q$  implies  ${\rm Im}({\widehat P}_{ii}),{\rm
		Im}({\widehat Q}_{ii})=0$.
	
	From the form of the RGEs one can already draw the following conclusions:
	
	\begin{enumerate}
		\item The lighter right-handed neutrinos can receive sizable contributions to their mass from the heaviest ones. This phenomenon can be traced back to the fact that (in general) the Lagrangian Eq.~(\ref{eq:Lagrangian}) possesses no global symmetry protecting the lighter right-handed masses \cite{Ibarra:2018dib,Aparici:2012vx}. Therefore, lighter right-handed neutrinos can receive sizable (or even dominant) contributions to their masses from quantum effects induced by heavier right-handed neutrinos. This effect, potentially very important for correctly describing the phenomenology of the seesaw model, is not manifest in a one-loop calculation and requires at least a two-loop calculation. 
		\item When the right-handed neutrino mass $M_i$ vanishes at the cut-off scale (or more generically, when the quantum contribution to the physical right-handed mass dominates over the tree-level contribution), the RGE evolution drives the right-handed mixing matrix to a quasi-fixed point in the infrared ({\it cf.} Eq.~(\ref{eq:Im_diag})):
		\begin{align}
		\sum_{k\neq i}M_k{\rm Im}\left(\hatP_{ki}^2\right)\simeq 0.
		\end{align}
		\item When two right-handed neutrino masses are degenerate at the cut-off scale (or more generically, when the quantum contribution to the physical mass difference between the two right-handed neutrinos dominates over the tree-level mass difference), the RGE evolution drives the right-handed mixing matrix to a quasi-fixed point in the infrared satisfying
		\begin{align}
		\sum_{k\neq i,j} M_k \hatP_{ki}\hatP_{kj}\simeq 0,
		\end{align}
		if  $M_i=M_j=0$ and $M_k\neq 0$ ({\it cf.} Eqs.~(\ref{eq:Re_off_diag}), (\ref{eq:Im_off_diag})), and 
		\begin{align}
			M_i{\rm Re}\left(\hatQ_{ij}\right)+2\sum_{k}M_k{\rm Re}\left(\hatP_{ki}\hatP_{kj}\right)\simeq 0 ,
		\end{align}
		if $M_i=M_j\neq 0$  ({\it cf.} Eq.~(\ref{eq:Re_off_diag})).
	\end{enumerate}
	
	Let us analyze in the following some implications of the two-loop quantum effects on seesaw models with two or three right-handed neutrinos, focusing on the case where the mass hierarchy is large at the cut-off scale, such that the physical mass of at least one of the right-handed neutrinos is dominated by the quantum effects induced by the heavier right-handed neutrinos. 
	
	\section{Two right-handed neutrino model}
	\label{sec:3}
	
	We consider first a simplified scenario containing only two right-handed neutrinos. We choose to work in the basis where the right-handed neutrino mass
	matrix at the cut-off scale $\Lambda$ is diagonal and real, with eigenvalues $M_1$ and $M_2$ (with $M_2> M_1$):
	\begin{align}
	M(\Lambda)=\begin{pmatrix} M_1 & 0 \\ 0 & M_2 \end{pmatrix}.
	\end{align}
	
	It is straightforward to integrate Eq.~(\ref{eq:RGE-mass-matrix}), to obtain the right-handed neutrino mass at the scale $\mu<\Lambda$ 
	(for details, see Appendix \ref{app:Picard}):
	\begin{align}
	M(t)\simeq  M +t \Big(M P + P^T M\Big) -\frac{t}{4}(1-2t)\Big(M P P+ P^T P^T M\Big)+t(4+t) P^T M P \;,
	\label{eq:Picard-2nd_order}
	\end{align}
	with $t\equiv \log(\mu/\Lambda)$.
	Since we are interested in the leading effects, we have neglected terms proportional to  $\mathcal{G}$ in Eq.~(\ref{eq:Q}), and we have kept terms up to ${\cal O}(P^2)$ which, as we will see below, can be crucial in some instances. 
	
	Below the scale $\mu\simeq M_2$ the phenomenology of the model can be conveniently described by the effective Lagrangian:
	\begin{align}
	{\cal L}_\mathrm{eff}\simeq  \frac{1}{2} \frac{Y_{\alpha 2} Y_{\beta 2}}{M_{22}}\left(\overline{L_\alpha} \widetilde{H} \right)\left(\widetilde{H}^T L^c_\beta\right)
	-\mathbb{Y}_{\alpha 1} \overline{L_\alpha } \widetilde H N_1
	-\frac{1}{2}\mathbb{M}_{11} \overline{N^c_1} N_1 + \mathrm{h.c.},
	\end{align}
	where the parameters must be evaluated at the scale $\mu\simeq M_2$ and where we have defined 
	\begin{align}
	\mathbb{Y}_{\alpha 1}&= Y_{\alpha 1} -\frac{M_{1 2} Y_{\alpha  2}}{M_{22}}\,, \nonumber \\
	 \mathbb{M}_{11}&=M_{11}- \frac{M_{1 2}M_{21}}{M_{22}}\,.
	 \end{align}
	We obtain for the physical mass of $N_1$ 
	\begin{align}
	M_{1}^{\rm phys}\simeq \mathbb{M}_{11}\Big|_{\mu=M_1}\simeq M_1-4 M_2 P_{21}^2 \log\left(\frac{\Lambda}{M_2}\right).
	 \label{eq:2loop1}
	\end{align}
	This result could have been also derived from Eq.~(\ref{eq:Re_diag}), using that at the cut-off scale the mixing matrix is $U_M=1$, and the fact that in this case $U_M$ does not change substantially under the RG running.
	
	Finally, for energy scales $\mu\lesssim M_{1}^{\rm phys}$ the lightest right-handed neutrino $N_1$ can also be integrated out, and the theory is simply described by the Weinberg operator
	\begin{align}
	{\cal L}_\mathrm{eff}\simeq \frac{1}{2} \left(\frac{Y_{\alpha 2}Y_{\beta 2}}{M_{22}}\Big|_{\mu=M_2}+
	\frac{\mathbb{Y}_{\alpha 1}\mathbb{Y}_{\beta 1}}{\mathbb{M}_{11}}\Big|_{\mu=M_1}\right)\left(\overline{L_\alpha} \widetilde{H} \right)\left(\widetilde{H}^T L^c_\beta\right)+ \mathrm{h.c.},
	\label{eq:2RHN-Weinberg}
	\end{align}
	from where one can calculate the active neutrino masses, running the Weinberg operator down to the electroweak scale \cite{Babu:1993qv}.\footnote{A numerical factor in the RGE was corrected in \cite{Antusch:2001ck}.} On the other hand, the most relevant RGE effects occur in this scenario at scales $\mu>M_2$, therefore one can approximate the neutrino mass eigenvalues by considering the values of the parameters frozen at $\mu=M_2$. Namely,
	\begin{align}
	{\cal M_\nu}\simeq \Big( Y M^{-1} Y^T\Big)\Big|_{\mu=M_2}\langle H^0\rangle^2.
	\label{eq:seesaw-formula}
	\end{align} 
	
	The most notable feature of the two-loop quantum effects is the contribution proportional to $M_2$ to the mass of the lightest right-handed neutrino, ({\it cf.} Eq.~(\ref{eq:2loop1})), which arises from the term $P^T M P$ in the RGE. This term may dominate over the tree-level contribution when the mass hierarchy is strong at the cut-off scale, and ought to be included for correctly describing the phenomenology of the model. 
		
	To emphasize the main implications of this term, we will consider in what follows the limit $M_1=0$, although the same conclusions hold as long as $M_1\ll 4 M_2 P_{21}^2 \log\left(\frac{\Lambda}{M_2}\right)$. 	
	In the toy model with only one lepton doublet $L_1$, discussed in
	Ref.~\cite{Ibarra:2018dib},  the active neutrino mass reads
	\begin{align}
	m_\nu&\simeq \left(\frac{Y_{12}^2}{M_{2}}\Big|_{\mu=M_2}+\frac{Y_{11}^2}{M_1}\Big|_{\mu=M_1}\right) \langle H^0\rangle^2\nonumber \\
	&\simeq  \left(Y_{12}^2-\frac{(16\pi^2)^2}{4Y_{12}^2 \log(\Lambda/M_2)}\right) \frac{\langle H^0\rangle^2}{M_2}, 
	\end{align}
	where we have used  $\mathbb{Y}_{11}\simeq Y_{11}$, which follows from the fact that 
	$M_{12}/M_{22} Y_{12}\sim Y_{11}Y^2_{12}/(16\pi^2) \ll Y_{11}$.   
	In this scenario, therefore, the active neutrino mass is mostly generated by the coupling of the lepton doublet to $N_1$. Yet, the neutrino mass is fairly insensitive to the value of $Y_{11}$, since the right-handed mass $M_1$ is generated by two-loop quantum effects induced by $N_2$, and depends itself on $Y_{11}$. More concretely, 
	\begin{align}
	|m_\nu|
	&  \simeq 0.05\, {\rm eV}\,\, \left(\frac{Y_{12}}{0.6}\right)^{-2} \left(\frac{M_2}{1.2\times 10^{19}\,{\rm GeV}}\right)^{-1}\;,
	\label{eq:nu-mass}
	\end{align}
	where we have assumed $M_2$ not far from the cut-off, such that  $\log(\Lambda/M_2)\simeq 1$. One then concludes that if the  right-handed neutrino mass spectrum is very hierarchical at the cut-off of the theory (which we assume to be the Planck scale), and if the lepton number is broken by a Majorana mass close to the cut-off scale, then quantum effects induced by a Yukawa coupling $Y_{12}={\cal O}(1)$  generate an active neutrino mass in the ballpark of the experimental values. 
	
	In the realistic case with three lepton doublets, the active neutrino masses can be obtained constructing the matrix invariants involving the neutrino mass matrix  Eq.~(\ref{eq:seesaw-formula}). Using the Faddeev--LeVerrier algorithm, and assuming ${\cal M}_\nu$ real, these are:
	\begin{align}
	I_1 =&~ \textrm{Tr}\left[\mathcal{M}_\nu\right] = (16\pi^2)\textrm{Tr}\left[M^{-1} P\right]\langle H^0\rangle^2, \nonumber\\
	I_2 =&~ 
	\frac{1}{2}\left(\textrm{Tr}\left[\mathcal{M}_\nu\right]^2 - \textrm{Tr}\left[\mathcal{M}_\nu^2\right]\right) =
	\frac{(16\pi^2)^2}{2}\left(\textrm{Tr}\left[M^{-1} P\right]^2 - \textrm{Tr}\left[M^{-1} PM^{-1} P\right]\right) \langle H^0\rangle^4,
	 \nonumber\\ 
	I_3 =&~\frac{1}{6}\Big(\textrm{Tr}\left[\mathcal{M}_\nu\right]^3 - 3\textrm{Tr}\left[\mathcal{M}_\nu\right]\textrm{Tr}\left[\mathcal{M}_\nu^2\right]+2\textrm{Tr}\left[\mathcal{M}^3_\nu\right]\Big)= \frac{(16\pi^2)^3}{6}\Big(\textrm{Tr}\left[M^{-1} P\right]^3\nonumber\\ & - 3\textrm{Tr}\left[M^{-1} P\right]\textrm{Tr}\left[ M^{-1}P M^{-1}P\right]+2\textrm{Tr}\left[M^{-1} PM^{-1} PM^{-1} P\right]\Big) \langle H^0\rangle^6.
	\label{eq:invariants_active}
	\end{align}	
		
\begin{figure}[t]
		\begin{center}
			\includegraphics[scale=0.35]{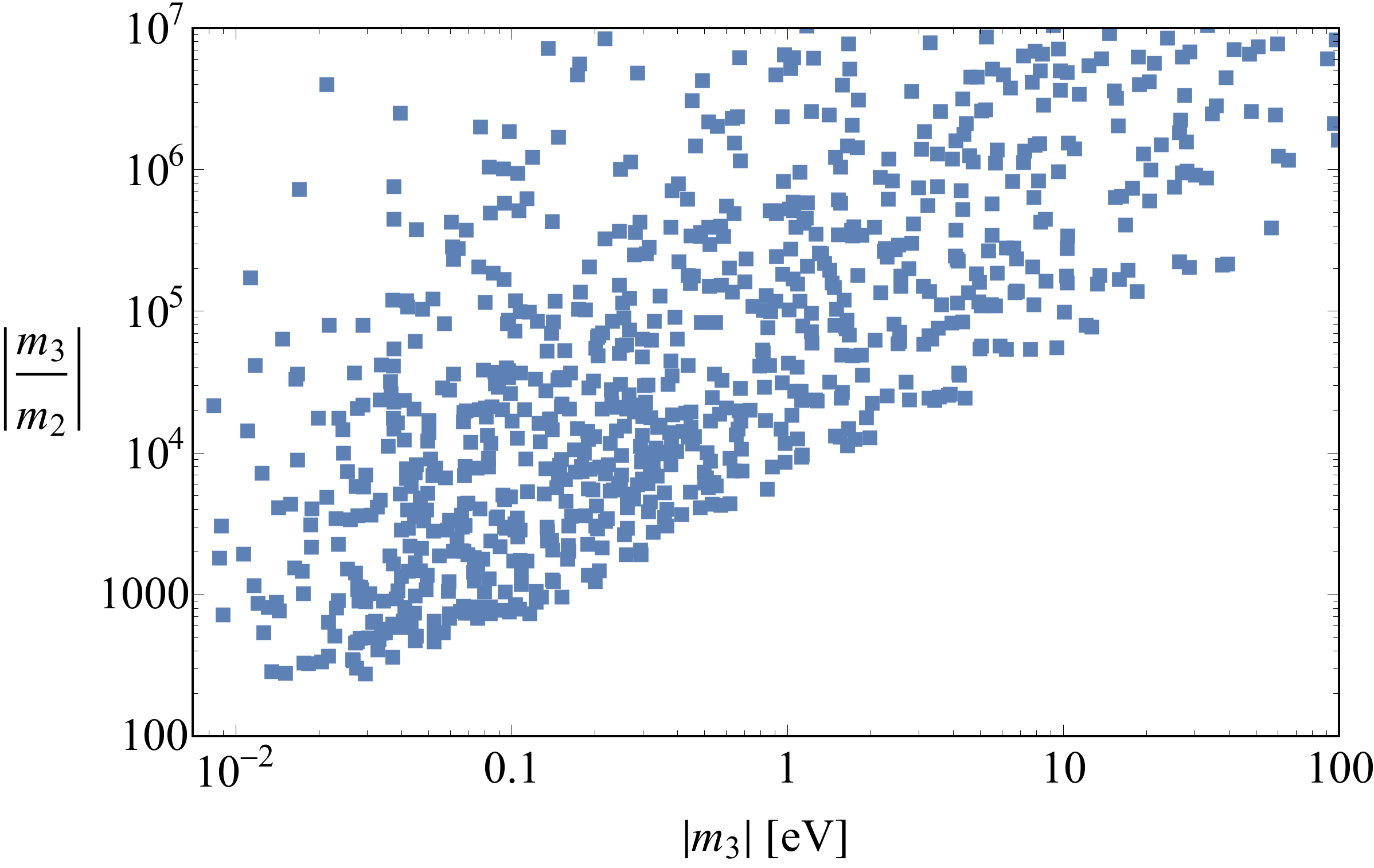}
			\caption{Scan plot with the predicted largest active neutrino mass $|m_3|$ and mass hierarchy $|m_3/m_2|$ for a model with two-right handed neutrinos and three lepton doublets. We have assumed $M_2=M_{\rm P}/\sqrt{8\pi}$, $M_1 = 0$ at the cut-off scale $\Lambda = M_{\rm P}$, and we have scanned the Yukawa eigenvalues in the range $10^{-2} \leq y_2 < y_1 \leq \sqrt{4 \pi}$ and the right-handed mixing angle between 0 and $2\pi$.}
			\label{fig:2RHN-model}
		\end{center} 
	\end{figure}

	For a model with three lepton doublets and two right-handed neutrinos, $I_3=0$. On the other hand, using that $I_3=m_1 m_2 m_3$ we obtain the well known result $m_1=0$.\footnote{Two loop quantum effects between the scale $M_1$ and the electroweak scale generate a non-vanishing value for $m_1$, which is however too small to be of phenomenological interest ~\cite{Petcov:1984nz,Davidson:2006tg}.} From the other two invariants, and keeping the leading terms in $P$, we obtain:\footnote{For the general complex case, the invariants must be constructed using the hermitian matrix ${\cal M}_\nu^\dagger {\cal M}_\nu$. Namely, $I_1={\rm Tr}[{\cal M}_\nu^\dagger{\cal M}_\nu]=m_2^2+m_3^2$, etc.}
	\begin{align}
	I_1 &= m_2+ m_3 \simeq (16\pi^2)\langle H^0 \rangle^2 \frac{P_{11}}{M_1}, \nonumber \\
	I_2 &= m_2 m_3\simeq  (16\pi^2)^2 \langle H^0 \rangle^4 \frac{P_{11} P_{22} - P_{12}^2}{M_1 M_2}.
	\end{align}
	Therefore, for the phenomenologically interesting case where the light neutrinos are hierarchical,
	\begin{align}
	m_3 &\simeq I_1\simeq (16\pi^2)\langle H^0 \rangle^2 \frac{P_{11}}{M_1}\Big|_{\mu=M_2},\nonumber\\
	m_2 &\simeq \frac{I_2}{I_1}\simeq (16\pi^2)\frac{\langle H^0 \rangle^2}{M_2} \Big(\frac{P_{11} P_{22} - P_{12}^2}{P_{11}}\Big)\Big|_{\mu=M_2}.
	\end{align}
	We parametrize the $P$-matrix as $P=\frac{1}{16\pi^2} U^T {\rm diag}(y_1^2,y_2^2) U$, with $U$ a $2\times 2$ unitary matrix and $y_1$, $y_2$ the Yukawa eigenvalues. Then, the light neutrino eigenvalues can be cast as:
	\begin{align}
	m_{3}&\simeq
	\frac{(16\pi^2)^2
		(y_1^2 U_{11}^2 +y_2^2 U_{21}^2 )\langle H^0\rangle^2}{4M_2  (y_2^2-y_1^2)^2 U^2_{11}U_{12}^2\log(M_2/\Lambda)}, \nonumber \\
	m_{2}& \simeq
	\frac{y_1^2 y_2^2\langle
		H^0\rangle^2 }{M_2 (y_1^2 U_{11}^2+y_2^2 U_{21}^2)},
	\end{align}
	which clearly only depend on right-handed mixing angles (see also \cite{Casas:2006hf}). Here, we have assumed that $M_2$ is not far from the cut-off scale, so that one can approximate $\log(\Lambda/M_2)\sim 1$. As in the one generation case, $m_3$ is in the ballpark of the experimental values, provided the right-handed mixing is sizable, and provided $y_2\sim 1$. However, this scenario tends to generate a mass hierarchy between $m_3$ and $m_2$ which is too large to explain oscillation experiments. Concretely, it can be checked that the mass hierarchy is bounded from below by:
	\begin{align}
	\Big|\frac{m_3}{m_2}\Big|\gtrsim \frac{(16\pi^2)^2}{(y_2^2 -y_1^2)^2\log(M_2/\Lambda)},
	\end{align}
	thus requiring fairly large Yukawa couplings to reproduce the experimental upper limit $|m_3/m_2|\lesssim 5.7$. The heaviest active neutrino mass and  mass hierarchy are shown in Fig.~\ref{fig:2RHN-model}, taking for concreteness $\Lambda=M_{\rm P} $, $M_2=M_{\rm P}/\sqrt{8\pi}$, and scanning the Yukawa eigenvalues in the range 
	$10^{-2} \leq y_2 < y_1 \leq \sqrt{4 \pi}$ and the right-handed mixing angle between 0 and $2\pi$. As seen for the Figure, it is possible to reproduce $m_3={\cal O}(0.05)$~eV, for appropriate choices of parameters, but the mass hierarchy tends to be too large. 
		
	\section{Three right-handed neutrino model}
	\label{sec:4}
	
	Finally, we consider the model with three right-handed neutrinos. At the cut-off scale $\Lambda$, the right-handed neutrino mass matrix is
	given by
	\begin{align}
	M(\Lambda)=\begin{pmatrix} M_1 & 0& 0\\0 & M_2 & 0 \\ 0& 0& M_3 \end{pmatrix},
	\end{align}
	with $M_1\ll M_2\ll M_3$. The mass $M_3$ receives
	typically small corrections, while $M_1$ and $M_2$ may receive sizable or
	even dominant contributions from quantum effects proportional to
	$M_3$. In either case, one can integrate $N_3$ out, leading to the effective Lagrangian 
	\begin{align}
	{\cal L}_\mathrm{eff}\simeq \frac{1}{2}\frac{Y_{\alpha 3}Y_{\beta 3} }{M_{33}}
	\left(\overline{L_\alpha } \widetilde{H} \right)\left(\widetilde{H}^T L^c_\beta\right)
	- \mathbb{Y}_{\alpha i}\overline{L_\alpha } \widetilde H N_i
	-\frac{1}{2} \mathbb{M}_{ij} \overline{N^c_i} N_j + \mathrm{h.c.},
	\label{eq:RHN3_1}
	\end{align} where we have defined
	\begin{align}
	\mathbb{Y}_{\alpha i}&= \Big(Y_{\alpha i} -\frac{M_{i 3}
		Y_{\alpha 3}}{M_{33}}  \Big),\nonumber \\
	\mathbb{M}_{ij}&=\Big(M_{ij}- \frac{M_{i 3}M_{j 3}}{M_{33}}\Big),
	\label{eq:BB}
	\end{align}
	with $i,j=1,2$ and couplings evaluated at the scale $\mu=M_3$. The Weinberg operator in Eq.~(\ref{eq:RHN3_1}) gives a contribution to the active neutrino masses $\lesssim
	10^{-6}$ eV. Thus, to study the generation of the atmospheric and solar mass scales it is sufficient to consider the role of $N_2$ and $N_1$. 
	The effective Lagrangian can then be  approximated by:
	\begin{align}
	{\cal L}_\mathrm{eff}\simeq -\mathbb{Y}_{\alpha i} \overline{L_\alpha} \widetilde H N_i -
	\frac{1}{2}\mathbb{M}_{ij} \overline{N^c_i} N_j+ \mathrm{h.c.}
	\label{eq:Lagrangian-2RHN}
	\end{align}	
	
	The right-handed masses at the scale $\mu=M_3$ can be calculated from the invariants
	\begin{align}
	I_1 &= \textrm{Tr}\left[\mathbb{M}\right] =M_1(t) + M_2(t) \simeq M_1+M_2+4M_2 P_{21}^2 t+4 M_3 (P_{31}^2 + P_{32}^2) t, \nonumber\\
	I_2 &= \textrm{det}\left[\mathbb{M}\right] =M_1(t) M_2(t) \simeq   M_1 M_2 + 4\left(M_1^2 P_{21}^2 + M_2^2 P_{21}^2\right) t  +4 M_3 \left(M_1 P_{32}^2 + M_2 P_{31}^2\right) t \nonumber \\
	&\hspace{2.3cm}+32 M_3^2 [P_{21}\left(P_{31}^2-P_{32}^2\right)-\left(P_{11}-P_{22}\right)P_{31}P_{32}]^2 t^3,
	 \label{eq:eigenvalues_eff_2RHN2}
	\end{align}
	with $t \equiv \mathrm{log}(\mu/\Lambda)$. In these expressions, we have kept only the leading contributions (including terms ${\cal O}(P^4)$ in the Picard expansion that, as we will see later, are relevant in some scenarios).
	In the generically expected case that the right-handed masses are hierarchical, one can approximate:
	\begin{align}
	\label{eq:eigenvalues_eff_2RHN}
	M_2\Big|_{\mu=M_3} &\simeq I_1, \nonumber \\ 
	M_1\Big|_{\mu=M_3} &\simeq \frac{I_2}{I_1}.
	\end{align}
	
	The masses of the right-handed neutrinos, and accordingly of the active neutrinos, crucially depend on the relative size of the quantum and the tree-level contributions to the right-handed neutrino masses. One can identify the following scenarios:
	\begin{enumerate}
		\item[{\it a})] Both right-handed neutrino masses are dominated by their tree-level values.
		\item[{\it b})] One right-handed neutrino mass is dominated by quantum contributions, while the other is dominated by its tree-level value.
		\item[{\it c})] Both right-handed neutrino masses are dominated by quantum contributions.
	\end{enumerate}
	Case ${\it a})$ corresponds to the well studied case in the literature, save for small quantum corrections that do not affect qualitatively the phenomenology. In the next subsections we will then focus on cases  ${\it b})$ and  ${\it c})$, where the predictions of the model can be significantly different to those from the tree-level (and even one-loop) calculations. Furthermore, since one (or two) tree-level parameters are washed-out by the RG evolution, the predictivity of the model gets enhanced. 
	
	\subsection{One right-handed neutrino mass dominated by quantum effects}
	
	We consider first the case where one of the tree-level masses is much larger than the radiative contributions proportional to $M_3$, while the other is much smaller. 
	Using Eqs.~(\ref{eq:eigenvalues_eff_2RHN2}) and (\ref{eq:eigenvalues_eff_2RHN}), we obtain for the masses at the scale $\mu=M_3$:
	\begin{align}
	M_i\Big|_{\mu=M_3}&\simeq 4 M_3 (P_{31}^2+P_{32}^2) \log\left(\frac{M_3}{\Lambda}\right),\nonumber \\
	M_j\Big|_{\mu=M_3}&\simeq   \frac{M_1 P_{32}^2+ M_2 P_{31}^2}{P_{31}^2+P_{32}^2},
	\label{eq:3RHN1}
	\end{align}
	where we have implicitly assumed that $M_i|_{\mu=M_3}$, $M_j|_{\mu=M_3}$ are much larger than the Dirac neutrino mass; this is generically the case for the Planck scale lepton number breaking scenario, and for generic structures in the Yukawa couplings. Note that the concrete mass ordering of these two right-handed neutrinos depends on the model parameters at the cut-off scale, hence we have left unspecified whether $i=1$ or 2 (and $j=2$ or 1). Note also that $M_j|_{\mu=M_3}$ is a linear combination of the two masses $M_1$ and $M_2$. This is due to the non-negligible right-handed mixing generated by the RG evolution. Concretely, at the scale $\mu=M_3$  the $2\times 2$ right-handed mass matrix in Eq.~(\ref{eq:Lagrangian-2RHN}) is diagonalized as $\mathbb{M}\simeq V {\rm diag}(M_i,M_j)V^T$ , with 
	\begin{equation}
	V\simeq\frac{1}{\sqrt{P_{31}^2+P_{32}^2}}\left(
	\begin{array}{cc}
	P_{32} & P_{31}\\
	-P_{31} & P_{32}
	\end{array}
	\right),
	\label{eq:V}
	\end{equation}
	which has in general sizable entries. 
	
	In the high-scale seesaw scenario under consideration in this paper, the heaviest right-handed neutrino (with mass $M_3\sim M_{\rm P}$) gives a contribution to the active neutrino masses $\lesssim 10^{-6}$ eV, and only the effects of the two lightest right-handed neutrinos are relevant for oscillation experiments. The active neutrino mass matrix can then be approximated using the two right-handed neutrino Lagrangian Eq.~(\ref{eq:Lagrangian-2RHN}). It reads:
	\begin{align}
	{\cal M}_\nu \simeq \mathbb{Y}\mathbb{M}^{-1}\mathbb{Y}^T\langle H^0\rangle^2\;.
	\end{align}
	
	The eigenvalues can be calculated along the lines of Section \ref{sec:3}, ({\it cf.} Eq.~(\ref{eq:invariants_active})): 
	\begin{align}
	I_1 &= \textrm{Tr}\left[\mathcal{M}_\nu\right] = (16\pi^2)\textrm{Tr}\left[\mathbb{M}^{-1} \mathbb{P}\right] \langle H^0\rangle^2,\nonumber\\
	I_2 &= 
	\frac{1}{2}\left(\textrm{Tr}\left[\mathcal{M}_\nu\right]^2 - \textrm{Tr}\left[\mathcal{M}_\nu^2\right]\right) =
	\frac{(16\pi^2)^2}{2}\left(\textrm{Tr}\left[\mathbb{M}^{-1} \mathbb{P}\right]^2 - \textrm{Tr}\left[\mathbb{M}^{-1} \mathbb{P}\mathbb{M}^{-1} \mathbb{P}\right]\right) \langle H^0\rangle^4,
	\nonumber\\ 
	I_3 &=\frac{1}{6}\Big(\textrm{Tr}\left[\mathcal{M}_\nu\right]^3 - 3\textrm{Tr}\left[\mathcal{M}_\nu\right]\textrm{Tr}\left[\mathcal{M}_\nu^2\right]+2\textrm{Tr}\left[\mathcal{M}^3_\nu\right]\Big)= \frac{(16\pi^2)^3}{6}\Big(\textrm{Tr}\left[\mathbb{M}^{-1} \mathbb{P}\right]^3\nonumber\\ & \hspace{0.3cm}- 3\textrm{Tr}\left[\mathbb{M}^{-1} \mathbb{P}\right]\textrm{Tr}\left[ \mathbb{M}^{-1} \mathbb{P}\mathbb{M}^{-1} \mathbb{P}\right]+2\textrm{Tr}\left[\mathbb{M}^{-1} \mathbb{P}\mathbb{M}^{-1} \mathbb{P}\mathbb{M}^{-1} \mathbb{P}\right]\Big) \langle H^0\rangle^6.
	\end{align}
	with $\mathbb{P}\equiv \frac{1}{16\pi^2}\mathbb{Y}^T\mathbb{Y}$.
		
	We obtain  $I_3=0$ , which implies $m_1=0$, and
	\begin{align}
	I_1 &= m_2+ m_3 \simeq \frac{16 \pi^2\langle H^0\rangle^2}{M_j|_{\mu=M_3}}\frac{P_{22}P_{31}^2+P_{11}P_{32}^2-2P_{31}P_{32}P_{21}}{P_{31}^2+P_{32}^2}\;, \nonumber \\
	I_2 &= m_2 m_3\simeq  \frac{\left(16 \pi^2\right)^2\langle H^0\rangle^4}{M_i|_{\mu=M_3} M_j|_{\mu=M_3}}\left(P_{11}P_{22}-P_{21}^2\right).
	\end{align}
	Therefore, for a hierarchical spectrum of light neutrinos, one obtains the following analytic expressions for $m_\alpha$ and $m_\beta$,

		\begin{align}
		m_\alpha&\simeq I_1 = \frac{16 \pi^2\langle H^0\rangle^2}{M_j|_{\mu=M_3}}\frac{P_{22}P_{31}^2+P_{11}P_{32}^2-2P_{31}P_{32}P_{21}}{P_{31}^2+P_{32}^2}\;, \nonumber \\
		m_\beta&\simeq \frac{I_2}{I_1}=
	\frac{16\pi^2\langle H^0\rangle^2}{M_i|_{\mu=M_3}}\frac{\left(P_{31}^2+P_{32}^2\right)\left(P_{11}P_{22}-P_{21}^2\right)}
	{P_{22}P_{31}^2+P_{11}P_{32}^2-2P_{31}P_{32}P_{21}},
	\end{align}
where we have left unspecified the labeling of the states, as the mass or ordering  is not determined a priori. Finally, using $P=\frac{1}{16\pi^2} U^T {\rm diag}(y_1^2, y_2^2, y_3^2) U$ we find that in most of the parameter space  the light neutrino masses can be approximated by
		\begin{align}
	m_\alpha&\simeq \frac{y_2^2\langle H^0\rangle^2}{M_2} \frac{U_{13}^2}{U_{31}^2} \;, \nonumber \\
	m_\beta&\simeq  \frac{(16 \pi^2)^2 \langle H^0\rangle^2}{4 M_3 y_3^2 U_{33}^2 \log(M_3/\Lambda)},
	\end{align}
	with all seesaw parameters now evaluated at the cut-off scale. Specifically, this expression corresponds to the generic case where $M_j|_{\mu=M_3}$ is dominated by the tree-level mass $M_2$. However, for special choices of parameters, namely when $P_{31}/P_{32}\ll \sqrt{M_1/M_2}$, $M_j|_{\mu=M_3}$ could be dominated by $M_1$. In this case, $m_\alpha\simeq  y_2^2 \langle H^0\rangle^2/M_1 \, U_{13}^2/U_{32}^2$.

\begin{figure}[t]
		\begin{center}
			\includegraphics[scale=0.63]{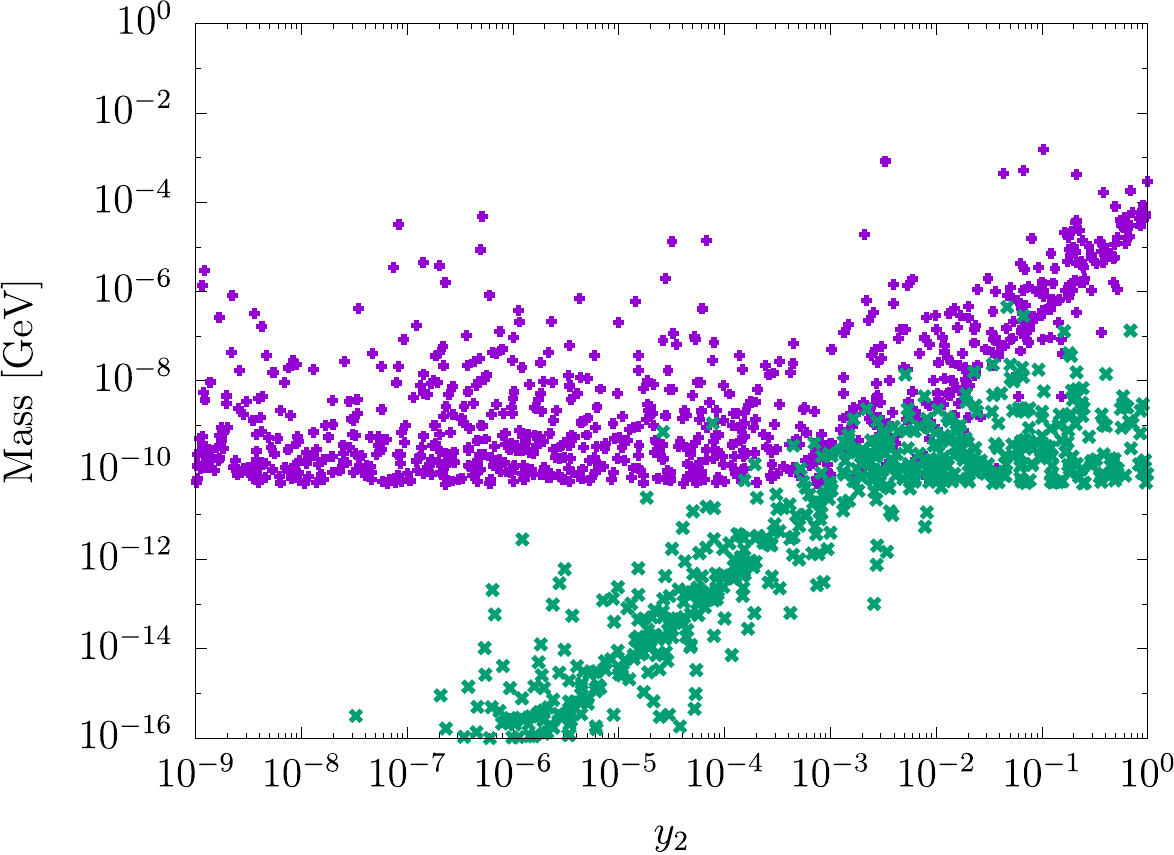}
			\includegraphics[scale=0.63]{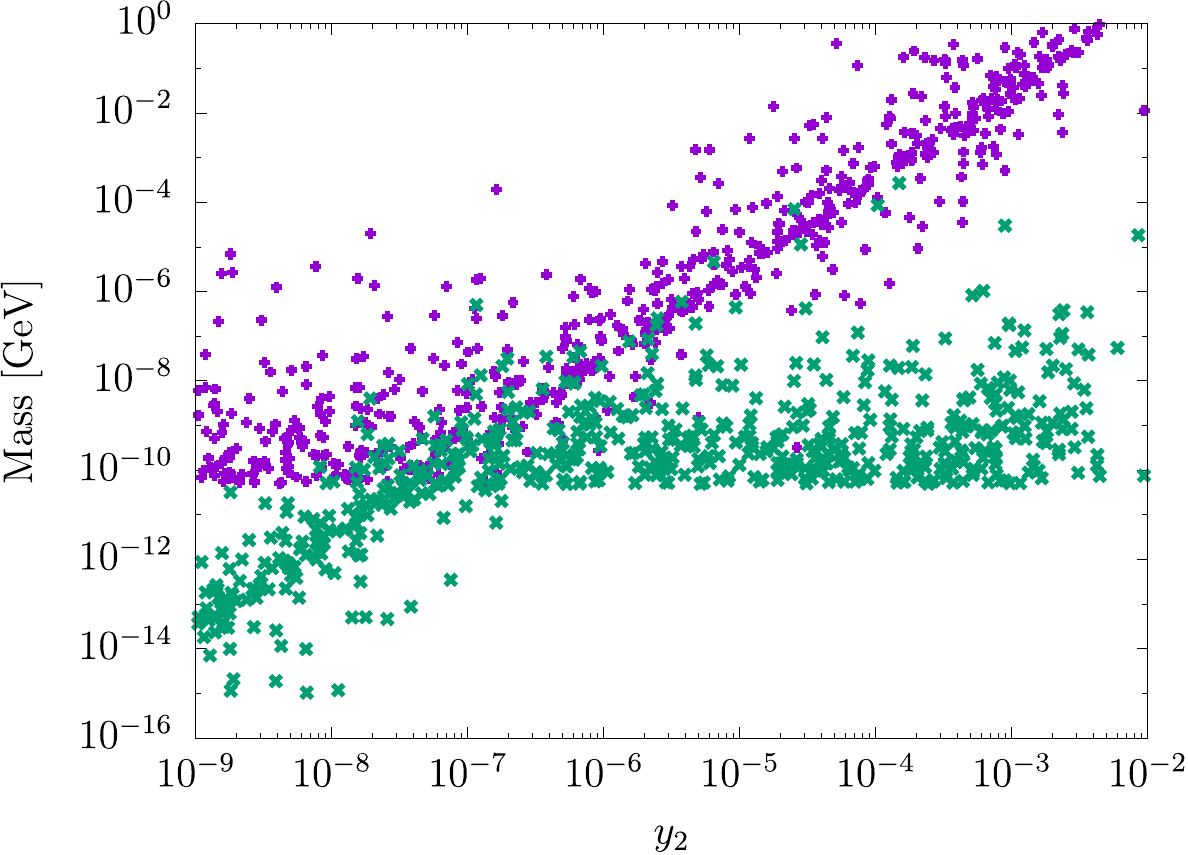}
			\caption{Scan plot of the active neutrino masses $|m_3|$ (magenta)  and  $|m_2|$
				(green) as a function of the  Yukawa eigenvalue $y_2$, for the scenario with $y_3=1$, $y_1=0$, $M_3=M_{\rm P}/\sqrt{8\pi}$, $M_2=10^9$ GeV (left plot) or $M_2=1$ GeV (right plot), and $M_1=0$ at the cut-off scale  $\Lambda=M_{\rm P}$. The right-handed mixing angles have been randomly scanned between 0 and $2\pi$. }
			\label{fig:2}
		\end{center}
	\end{figure}

	Alongside to our analysis for the two right-handed neutrino case analyzed in Section~\ref{sec:3}, we show in Fig.~\ref{fig:2} the expected masses $m_3$ (violet) and $m_2$ (green) as a function of the Yukawa eigenvalue $y_2$, resulting from a random scan of the mixing angles between 0 and $2\pi$, and assuming $y_3=1$, $y_1=0$, $\Lambda=M_{\rm P}$, $M_3=M_{\rm P}/\sqrt{8\pi}$ and $M_2=10^9$~GeV (left plot) or $M_2=1$~GeV (right plot). One of the mass eigenvalues is in the ballpark of the experimental values and can be adjusted by an appropriate choice of $y_3$, $M_3$ and/or the mixing angles. The other mass eigenvalue, on the other hand, is very sensitive to $y_2$ and $M_2$, and requires special choices of parameters (not necessarily fine-tuned). For example, for the parameters of the left (right) plot, it is necessary to postulate $y_2\sim 10^{-2}$ ($\sim 10^{-7}$) to reproduce the observations. 
	
	\subsection{Two right-handed neutrino masses dominated by quantum effects}
	
	Let us turn now to the case where the tree level right-handed masses $M_1$ and $M_2$  are much smaller than the quantum contributions proportional to $M_3$. The mass matrix at the scale $\mu=M_3$ can be diagonalized as $\mathbb{M}\simeq V {\rm diag}(M_1,M_2) V^T$, where $V$ is still given by Eq.~(\ref{eq:V}) and the eigenvalues are
	\begin{align}
	M_2\Big|_{\mu=M_3}&\simeq
	4M_3\left(P_{31}^2+P_{32}^2\right)\log\left(\frac{M_3}{\Lambda}\right)\;, \nonumber \\
	M_1\Big|_{\mu=M_3}&\simeq
	8M_3\frac{\Bigl(P_{21}\left(P_{31}^2-P_{32}^2\right)-\left(P_{11}-P_{22}\right)P_{31}P_{32}\Bigr)^2}{P_{31}^2+P_{32}^2}\log^2\left(\frac{M_3}{\Lambda}\right)\;,
	\label{eq:M1M2}
	\end{align}
	as follows from Eqs.~(\ref{eq:eigenvalues_eff_2RHN2}) and (\ref{eq:eigenvalues_eff_2RHN}). It can be checked that at order $P^2$ the mass matrix has one vanishing eigenvalue, however two non-vanishing eigenvalues are generically expected since there is no symmetry protecting the lightest eigenvalue (for generic Yukawa structures). The non-zero value for $M_1|_{\mu=M_3}$ arises at order $P^4$ as explicit in Eq.~(\ref{eq:M1M2}). Further, using the eigendecomposition $P=\frac{1}{16\pi^2}U^T{\rm diag}(y_1^2,y_2^2,y_3^2)U$ one finds, 
	\begin{align}
		M_2\Big|_{\mu=M_3}&\simeq 
		\frac{4}{(16\pi^2)^2}M_3 y_3^4 U_{33}^2(U_{31}^2+U_{32}^2)\log\left(\frac{M_3}{\Lambda}\right)\;,
	\nonumber \\
	M_1\Big|_{\mu=M_3}&\simeq\frac{8}{(16\pi^2)^4} M_3y_2^4 y_3^4 \frac{U^2_{13}U^2_{23}}{(U_{31}^2+U_{32}^2)}\log^2\left(\frac{M_3}{\Lambda}\right)\;,
	\label{eq:M1M2-U}
	\end{align}
	where we have assumed $y_1\ll y_2\ll y_3$. It is therefore necessary to have a Yukawa matrix of rank $\geq 2$ in order to generate radiatively a second mass. 	
	
	The largest eigenvalue is typically much larger than any Dirac neutrino mass and practically coincides with the physical mass of the eigenstate $N_2$, namely $M_2^{\rm phys}\simeq M_2\Big|_{\mu=M_3}$. On the other hand, the lightest eigenvalue $M_1|_{\mu=M_3}$ is suppressed by the large factor $(16\pi^2)^4$, as well as by the factor $y_2^4$. Therefore, even despite being proportional to the large mass $M_3$, the resulting eigenvalue can be small and the corresponding physical state could have a sizable mixing with the active neutrino states. 	
	
	It is convenient to work in the basis of right-handed neutrinos where the mass matrix $\mathbb{M}$ is diagonal. Defining $N'=V^T N$ one obtains
	\begin{align}
	{\cal L}_\mathrm{eff}\simeq -\mathbb{Y}^\prime_{\alpha i} \overline{L_\alpha} \widetilde H N^\prime_i -
	\frac{1}{2}M_{i} \overline{N^{\prime c}_i} N^\prime_i+ \mathrm{h.c.},
	\end{align}
	where $\mathbb{Y}^\prime = \mathbb{Y} V$. After integrating out $N'_2$ we obtain
	\begin{align}
	{\cal L}_\mathrm{eff}\simeq \frac{1}{2}\frac{\mathbb{Y}_{\alpha 2}^{\prime}\mathbb{Y}_{\beta 2}^{\prime} }{M_{2}}\left(\overline{L_\alpha} \widetilde{H} \right)\left(\widetilde{H}^T L^c_\beta\right)
	-\mathbb{Y}_{\alpha1}^{\prime} \overline{L_\alpha} \widetilde H N_1^\prime
	-\frac{1}{2}M_{1} \overline{N_1^{\prime c}} N_1^\prime + \mathrm{h.c.}
	\end{align}
	This Lagrangian yields after electroweak symmetry breaking the following $4 \times 4$ neutrino mass matrix 
	\begin{align}
	\mathcal{M}_\nu \simeq \begin{pmatrix} - \frac{\mathbb{Y}_{\alpha 2}^{\prime}\mathbb{Y}_{\beta 2}^{\prime} }{M_{2}}\langle H^0\rangle^2 &  \mathbb{Y}^\prime_{\alpha 1} \langle H^0\rangle\\  (\mathbb{Y}^{\prime}_{\alpha 1})^T\langle H^0\rangle & M_1 \end{pmatrix}.
	\label{eq:4x4matrix}
	\end{align}
	To calculate the eigenvalues, 
	it is useful to decompose the mass matrix as:
	\begin{align}
	\mathcal{M}_\nu \simeq \textfrak{m}_1 u_1 u_1^T + \textfrak{m}_2 u_2 u_2^T + \textfrak{m}_3 \left(u_3 u_3^T - u_4 u_4^T\right),
	\label{eq:matrix-mus}
	\end{align}
	where $u_i$ are the vectors
	\begin{align}
	u_1 = \frac{1}{\sqrt{\mathbb{P}_{22}^\prime}}\begin{pmatrix} \mathbb{Y}_{12}^\prime \\ \mathbb{Y}_{22}^\prime \\ \mathbb{Y}_{32}^\prime \\ 0 \end{pmatrix}, \qquad u_2 = \begin{pmatrix} 0 \\ 0 \\ 0 \\ 1\end{pmatrix}, \qquad u_{3,4} = \frac{1}{\sqrt{2 \mathbb{P}_{11}^\prime}} \begin{pmatrix} \mathbb{Y}_{11}^\prime \\ \mathbb{Y}_{21}^\prime \\ \mathbb{Y}_{31}^\prime \\  \pm \sqrt{\mathbb{P}_{11}^\prime},\end{pmatrix},
	\end{align}
	which are normalized to unity (but not forming an orthonormal set), and $\textfrak{m}_i$ are mass scales (defined such that $\textfrak{m}_i\geq 0$) given by:
	\begin{align}
	\textfrak{m}_1 = -\frac{\mathbb{P}^\prime_{22}}{M_2} \langle H^0\rangle^2, \qquad \textfrak{m}_2 = M_1, \qquad \textfrak{m}_3 = \sqrt{\mathbb{P}_{11}^\prime} \langle H^0\rangle.
	\label{eq:mus}
	\end{align}
	Here, we have introduced the $2\times 2$ matrix $\mathbb{P}^\prime = \mathbb{Y}^{\prime \dagger} \mathbb{Y}^\prime = V^\dagger \mathbb{Y}^\dagger \mathbb{Y} V$, with $\mathbb{Y}$ and $V$ respectively given in Eq.~(\ref{eq:BB}) and Eq.~(\ref{eq:V}). Explicitly, the matrix elements read:
	\begin{align}
	\mathbb{P}_{11}^\prime &\simeq (16\pi^2)\frac{P_{22} P_{31}^2 + P_{11} P_{32}^2 - 2 P_{12} P_{31} P_{32}}{P_{31}^2 + P_{32}^2} \;,\nonumber\\
	\mathbb{P}_{22}^\prime & \simeq (16\pi^2)\frac{P_{22} P_{32}^2 + P_{11} P_{31}^2 + 2 P_{12} P_{31} P_{32}}{P_{31}^2 + P_{32}^2}\;, \nonumber\\
	\mathbb{P}_{12}^\prime & \simeq (16\pi^2)\frac{P_{12}(P_{31}^2 + P_{32}^2) + P_{31} P_{32}(P_{11}-P_{22})}{P_{31}^2+P_{32}^2}\;.
	\end{align}
	Using the eigendecomposition $P=\frac{1}{16\pi^2}U^T{\rm diag}(y_1^2,y_2^2,y_3^2)U$ one finds, 
	\begin{align}
	 \mathbb{P}_{11}^\prime &\simeq \frac{U_{13}^2}{U_{31}^2+U_{32}^2} y_2^2 \;,\nonumber\\
	 \mathbb{P}_{22}^\prime &\simeq (U_{31}^2+U_{32}^2) y_3^2 \;,\nonumber \\
	 \mathbb{P}_{12}^\prime &\simeq \frac{2U_{31}^3 U_{32}}{U_{31}^2+U_{32}^2} y_3^2 \;.
	 \label{eq:Pbar}
	 \end{align}
	 
	 One can check from the mass matrix Eq.~(\ref{eq:matrix-mus}) that ${\rm det}({\cal M}_\nu)=0$, which implies that the lightest eigenvalue is massless, $m_1=0$. The remaining three eigenvalues ($m_3$, $m_2$ and $m_s$) can be calculated analytically solving a cubic equation. The resulting expression is complicated and provides little insight on the form of the solution. On the other hand, generically one among the three mass scales $\textfrak{m}_i$  will be much larger than the other two. In these cases, it is possible to derive (using degenerate perturbation theory) simple analytical expressions for the eigenvalues.  We can distinguish the following three cases:
	\begin{enumerate}
		\item[{\it i})] $\textfrak{m}_2 \gg \textfrak{m}_1, \textfrak{m}_3$.
		
		The non-vanishing eigenvalues read:
	\begin{align}
	m_s &\simeq \textfrak{m}_2 = M_1, \nonumber\\
	m_3 &\simeq -\frac{\textfrak{m}_3^2}{\textfrak{m}_2} = -\frac{\mathbb{P}_{11}^\prime}{M_1} \langle H^0\rangle^2, \nonumber\\
	m_2 &\simeq \textfrak{m}_1 = -\frac{\mathbb{P}_{22}^\prime}{M_2} \langle H^0\rangle^2. 
	\end{align}
	The heaviest eigenstate is an almost sterile neutrino with mass approximately equal to the lightest right-handed neutrino mass. The other two eigenstates are active, with small masses due to the seesaw mechanism. In terms of the cut-off parameters, the masses read:
		\begin{align}
		m_s&\simeq8 M_3\frac{ y_2^4 y_3^4}{(16 \pi^2)^4} \frac{U_{13}^2 U_{23}^2}{U_{31}^2 +U_{32}^2}
	\log^2\left(\frac{M_3}{\Lambda}\right),\nonumber \\
		m_3&\simeq-\frac{(16 \pi^2)^4 \langle H^0\rangle^2}{8 M_3 y_2^2 y_3^4 U_{23}^2\log^2\left(M_3/\Lambda\right) },\nonumber \\
m_2&\simeq-\frac{(16\pi^2)^2\langle H^0\rangle^2}{4 M_3 y_3^2 U_{33}^2\log\left(M_3/\Lambda\right)}.
	\end{align}
	
	Analogously to Eq.~(\ref{eq:nu-mass}), one finds for this scenario
	\begin{align}
	|m_2|
	&  \simeq 0.05\, {\rm eV}\,\, \left(\frac{y_3 U_{33}}{0.6}\right)^{-2} \left(\frac{M_3}{1.2\times 10^{19}\,{\rm GeV}}\right)^{-1}\;,
	\label{eq:nu-mass-3RHN}
	\end{align}
	and is in the ballpark of the experimental values for typical values of the parameters (here we have assumed $\log\left(\Lambda/M_3\right)\sim 1$). On the other hand, the mass hierarchy between $m_3$ and $m_2$ is
	\begin{align}
	\Big|\frac{m_3}{m_2}\Big|\simeq \frac{(16\pi^2)^2}{2 y_2^2 U_{23}^2 \log\left(\Lambda/M_3\right)},
	\end{align} 
	which is always much larger than the observed value.

	\item[{\it ii})] $\textfrak{m}_3 \gg \textfrak{m}_1, \textfrak{m}_2$.

The non-vanishing eigenvalues read:
\begin{align}
m_{\alpha,s} &\simeq \pm \textfrak{m}_3 = \pm  \sqrt{\mathbb{P}_{11}^\prime}\langle H^0\rangle, \nonumber \\
m_\beta &\simeq \frac{\mathbb{P}_{11}^\prime \mathbb{P}_{22}^\prime - \mathbb{P}_{12}^{\prime 2}}{\mathbb{P}_{22}^\prime} \frac{\textfrak{m}_1}{\textfrak{m}_3^2} \langle H^0\rangle^2,
\end{align}
where we have left unspecified the labeling of the states, as the mass ordering is not determined a priori. The sterile neutrino forms in this case a pseudo-Dirac state with one of the active neutrinos. 

	In terms of the cut-off parameters, the masses read:
	\begin{align}
	m_{\alpha,s}&\simeq \pm y_2 \langle H^0\rangle\sqrt{\frac{U_{13}^2}{U_{31}^2+U_{32}^2}}, \nonumber \\
	m_\beta&\simeq-\frac{(16 \pi^2)^2 \langle H^0\rangle^2}{4 M_3 y_3^2 U_{33}^2\log\left(M_3/\Lambda\right)}.
	\end{align}
	
	The mass $m_\beta$ is given by Eq.~(\ref{eq:nu-mass-3RHN}) and again lies in the ballpark of the experimental values. The mass hierarchy between $m_\alpha$ and $m_\beta$ reads:
	\begin{align}
	\Big|\frac{m_\alpha}{m_\beta}\Big|=\frac{M_3 y_2 y_3^2}{64\pi^4 \langle H^0\rangle} U_{33}^2\sqrt{\frac{U_{13}^2}{U_{31}^2+U_{32}^2}}\log\left(\frac{\Lambda}{M_3}\right),
	\end{align}
	and can be larger or smaller than one. In particular, it can take the observed value for appropriate parameters.

\item[{\it iii})] $\textfrak{m}_1 \gg \textfrak{m}_2, \textfrak{m}_3$.
	
	The non-vanishing eigenvalues read:
	\begin{align}
	m_\alpha &\simeq \textfrak{m}_1 =-\frac{\mathbb{P}_{22}^\prime}{M_2} \langle H^0\rangle ^2,\nonumber \\
	m_{s,\beta} &\simeq \frac{1}{2}\left(\textfrak{m}_2 \pm \sqrt{\textfrak{m}_2^2 + 4 \textfrak{m}_3^2 - 4 \frac{\mathbb{P}_{12}^{\prime 2}}{\mathbb{P}_{22}^\prime}\langle H^0\rangle ^2}\right),
	\end{align}
	where we have left unspecified the labeling of the states, as the mass ordering is not determined a priori. As in case {\it ii}), the sterile neutrino forms a pseudo-Dirac state with one of the active neutrinos.
	
	Using Eqs.~(\ref{eq:mus}) and (\ref{eq:Pbar}) one finds that $\textfrak{m}_3\sim y_2 \langle H^0\rangle$. Therefore, reproducing the correct neutrino parameters typically requires $y_2\lesssim 10^{-12}$, which implies $\textfrak{m}_3\gg\textfrak{m}_2=M_1$. The masses in terms of the cut-off parameters then take a simple form:
	\begin{align}
	m_\alpha&\simeq -\frac{(16 \pi^2)^2 \langle H^0\rangle^2}{4 M_3 y_3^2 U_{33}^2  \log\left(M_3/\Lambda\right)}, \nonumber \\
	m_{s,\beta}&\simeq \pm y_2
	\langle  H^0\rangle\sqrt{\frac{U_{13}^2}{U_{31}^2+U_{32}^2}},
	\label{eq:case_ii}
	\end{align}
	which leads to the same results as in case {\it ii}). 
	
	\end{enumerate}
	
\begin{figure}[t]
		\begin{center}
			\includegraphics[scale=0.63]{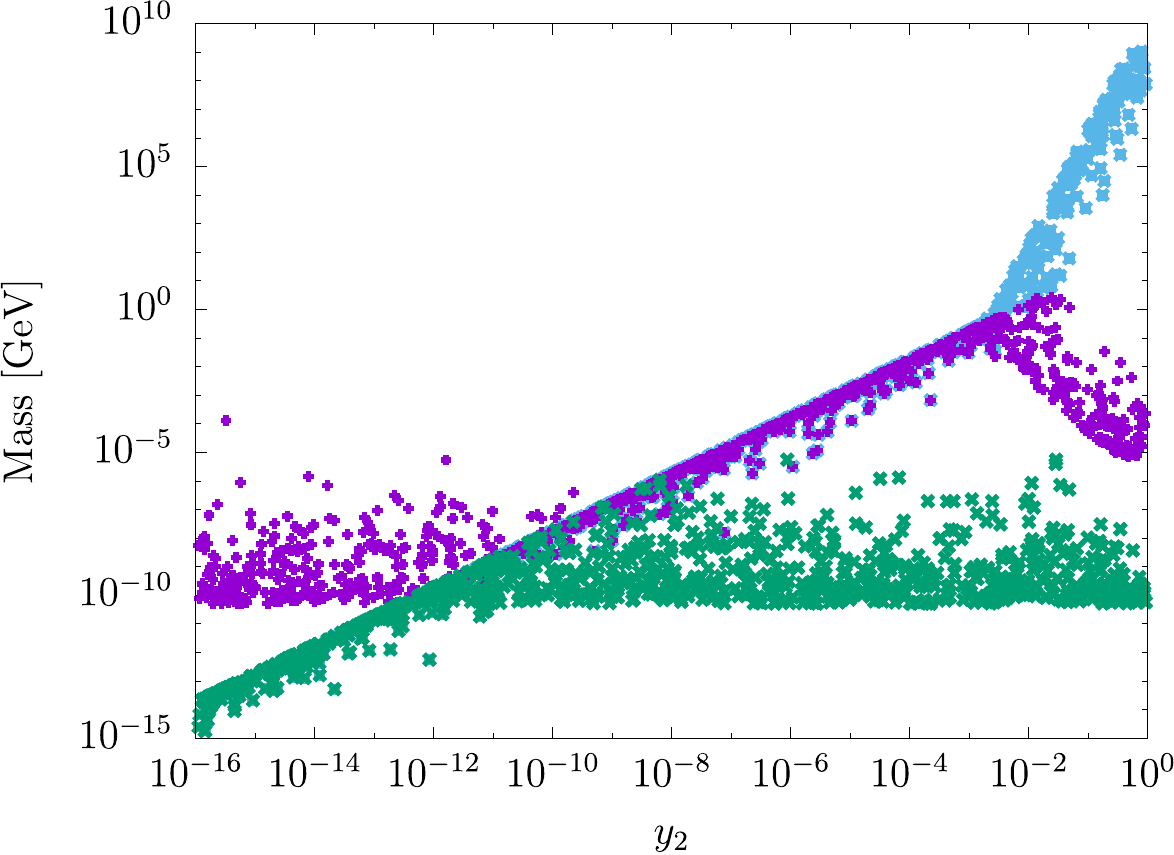}
			\caption{Scan plot of the active neutrino masses $|m_3|$ (magenta),  $|m_2|$
				(green), and the lightest sterile neutrino mass $|m_s|$ (blue) as a function of the  Yukawa eigenvalue $y_2$, for the scenario with $y_3=1$, $y_1=0$, $M_3=M_{\rm P}/\sqrt{8\pi}$ and $M_1=M_2=0$ at the cut-off scale  $\Lambda=M_{\rm P}$. The right-handed mixing angles have been randomly scanned between 0 and $2\pi$. 
}
			\label{fig:3}
		\end{center} 
	\end{figure}

	We show in Fig.~\ref{fig:3} a scan plot of the mass eigenvalues resulting from the diagonalization of the full $6\times 6$ mass matrix, as a function of $y_2$, after solving numerically the two-loop RGE equations for $y_1=0$, $y_3=1$, $\Lambda=M_{\rm P}$, $M_3=M_{\rm P}/\sqrt{8\pi}$ and random values of the mixing angles at the cut-off scale.  We only show for simplicity $m_s$, $m_3$ and $m_2$ to illustrate our discussion, and we omit $M_3$ and $M_2$, which are much heavier, or $m_1$ which is much lighter. 
	In the figure one can notice the regimes discussed above: case {\it i}) corresponds to 
	$y_2\gtrsim10^{-2}$, case {\it ii}) to 
	$10^{-12}\lesssim y_2 \lesssim 10^{-3}$ and case {\it iii}) to $y_2\lesssim10^{-12}$ respectively. 

	As generically expected, one of the eigenvalues lies in the ballpark of the experimental values. With our assumptions, case {\it i}) leads to a hierarchy between the atmospheric and the solar mass scales which is much larger than the measured value. Only when $y_2\sim 10^{-13}-10^{-11}$, corresponding to the region between cases {\it ii}) and {\it iii}), the mass hierarchy is in qualitative agreement with the measured value. In this case we expect one of the measured mass scales (solar or atmospheric) to be of Majorana type, and the other to be of Dirac type. 

	Let us finish stressing that for the plot we have assumed $y_3=1$, $\Lambda=M_{\rm P}$, $M_3=M_{\rm P}/\sqrt{8\pi}$  motivated by the fact that one of the predicted neutrino mass scales is close to the measured experimental value. However, the analysis could be extended to other cut-off parameters and there could be viable regions of the parameter space for case {\it i}), {\it ii}) or {\it iii}), or the intermediate regimes.
	
	\section{Conclusions}
	\label{sec:5}
	
	In a given model, quantum effects modify the values of the parameters at low energies compared to their values at the cut-off scale of the model. The differences can be significant when those parameters are not protected by symmetries. In this work we have investigated quantum effects on the right-handed neutrino parameters in the seesaw framework. Due to the violation of the total and family lepton numbers, there is no symmetry protecting the lighter right-handed masses against quantum corrections induced by the heavier right-handed neutrinos. As a consequence, quantum effects can generate sizable (or even dominant) contributions to the physical masses of the lighter right-handed neutrinos. The leading effect appears at the two-loop level. In this paper we have calculated the two-loop $\beta$-function for the right-handed Majorana neutrino mass matrix for the Standard Model extended with right-handed neutrinos, complementing the already existing two-loop $\beta$-functions for the rest of the parameters of the model, and thus completing the set of two-loop $\beta$-functions of the seesaw model.
	
	We have also analyzed some phenomenological implications of the two-loop renormalization group evolution. Concretely, we have studied scenarios where one of the right-handed neutrino masses (or two) are dominated by quantum effects, such that the values of the masses at the cut-off scale are washed out by the renormalization group evolution. For our numerical analysis we have focused in the case where the total lepton number is broken by a right-handed neutrino mass close to the Planck scale. This scenario is particularly predictive, as some of the physical masses of the right-handed neutrinos (and of the light neutrinos via the seesaw mechanism) are ultimately related to known scales. Furthermore, we have identified the quasi-fixed points in the infrared for the right-handed neutrino mixing matrix. 
	
	We have first analyzed the scenario with two-right handed neutrinos, assuming that the lightest mass is dominated by quantum effects induced by the heaviest mass, which we assume to be close to the Planck scale.	We find that one of the active neutrinos has a mass which lies, for ${\cal O}(1)$ Yukawa couplings, in the ballpark of the experimental values. However, this scenario tends to generate a mass hierarchy which is much larger than the one inferred from oscillation experiments. The correct mass hierarchy can be obtained by adding to the model a third right-handed neutrino, with an intermediate scale mass, larger than the quantum effects induced by the heaviest right-handed neutrino. In this scenario, one of the active neutrino masses lies in the ballpark of the experimental values; the second active mass is not predicted, but can be reproduced by adjusting the right-handed neutrino mass and the Yukawa couplings. Finally, we have considered a scenario with three right-handed neutrinos, where two of the masses are dominated by quantum effects. Again, we find that one of the active neutrino masses lies in the ballpark of the experimental values. The second active neutrino mass can be reproduced by adjusting the Yukawa couplings of the model. We find that this scenario generically leads to a Dirac mass, either for the active neutrino involved in atmospheric neutrino oscillations or for the one involved in solar neutrino oscillations, which could have implications for neutrinoless double beta decay experiments.

	\section*{Acknowledgements}
	This work has been supported by the Collaborative Research Center SFB1258, by the
	Deutsche Forschungsgemeinschaft (DFG, German Research Foundation) under Germany's Excellence
	Strategy -- EXC-2094 -- 390783311, and by the NSERC (Natural Sciences and Engineering Research Council of Canada). 
	The numerical computations in this work have been carried out at the Yukawa Institute Computer Facility.

	\appendix
	\section{Picard series}
	\label{app:Picard}
	The renormalization group equation of the right-handed mass matrix reads
	\begin{align}
	\frac{dM(t)}{dt}=\beta_M[M(t)],
	\end{align}
	where $t\equiv\log(\mu/\Lambda)$, and the $\beta$ function at one- and two-loops are given in Eq.~(\ref{eq:beta-2loop}). The solution to the RGE can be formally written as:
	\begin{align}
	M(t)=M_0+\int_0^t \beta_M[M(t')] dt',
	\end{align}
	with $M(t=0)\equiv M_0$. The solution  can be found iteratively, using a Picard series. For $n=0$ the solution is simply  $M^{(0)}(t)=M_0$, and for $n\geq 1$
	\begin{align}
	M^{(n)}(t)&=M_0+\int_0^t \beta_M[M^{(n-1)}(t')] dt'.
	\end{align} 
	Explicitly, the solution at first order and second order reads:
	\begin{align}
	M^{(1)}(t)&=M_0+\int_0^t \beta_M[M^{(0)}(t)] dt'=M_0+\beta_M[M_0] t,\\
	M^{(2)}(t)&=M_0+\int_0^t \beta_M[M^{(1)}(t)] dt'\nonumber\\
	&=M_0+\beta_M[M_0]t+\frac{1}{2}(\beta_M\circ\beta_M)[M_0] t^2,
	\end{align}
	where we have used that the $\beta$-function is linear in $M$, and we
	have denoted the function composition $(f\circ g)(x)=f(g(x))$. In
	general, the solution can be written as:
	\begin{align}
	M^{(n)}(t)=\sum_{k=0}^n \frac{1}{k!}\beta_M^k[M_0] t^k,
	\label{eq:Picard-general}
	\end{align}
	with $\beta_M^n=\underbrace{\beta_M\circ \beta_M\circ\dots \circ \beta_M}_{n-{\rm times}}$, and $\beta_M^0=1$.
	
		The general form of the $\beta$-function is:
		\begin{align}
		\beta_M= \sum_{n,m} a_{nm} \left(P^T\right)^n M P^m,
		\end{align}
		where $a_{nm}=a_{mn}$, and which are scalar functions in flavor space (and thus depending on gauge couplings and traces of Yukawa couplings). Substituting the $\beta$-function in the Picard expansion Eq.~(\ref{eq:Picard-general}) we obtain, up to ${\cal O}(P^4)$ terms,
		\begin{align}
		M(t)\simeq &~M  +a_{10} t \Big(M P + P^T M\Big)  \nonumber \\
		& +\frac{t}{2}\Big[a_{10}^2 t + 2a_{20}\Big]\Big(M P P+ P^T P^T M\Big)\nonumber\\
		&+t\Big[a_{11} + a_{10}^2 t \Big]P^T M P \nonumber\\
		&+\frac{t}{6}\Big[6 a_{30} +6 a_{10} a_{20} t + a_{10}^3 t^2\Big]\Big(M PPP+ P^T P^T P^T M\Big) \nonumber\\
		&+\frac{t}{2}\Big[2 a_{21} + 2 a_{10} (a_{11}+ a_{20}) t + a_{10}^3 t^2\Big]\Big(P^T M PP +P^T P^T M P\Big)\nonumber\\
		&+\frac{t}{24}\Big[24 a_{40} +  (12a_{20}^2  + 24 a_{10} a_{30}) t + 12 a_{10}^2 a_{20} t^2 +  a_{10}^4 t^3\Big]\Big(M PPPP+ P^T P^T P^T P^T M\Big)\nonumber \\
		&+\frac{t}{6}\Big[6 a_{31} + 6\Big( a_{10} (a_{21} + a_{30}) + a_{11} a_{20}\Big) t +3 a_{10}^2 (a_{11} +2 a_{20}) t^2 + a_{10}^4 t^3\Big]\nonumber \\ 
		&\times \Big( P^T M PPP+P^T P^T P^T M P\Big) \nonumber \\ 
		&+\frac{t}{4}\Big[4a_{22} + 2 (a_{11}^2  + 2 a_{20}^2  + 4 a_{10} a_{21})t + 4 a_{10}^2 (a_{11}  +  a_{20}) t^2 + a_{10}^4 t^3\Big]P^T P^T M P P.
		\label{eq:Picard-approx}
		\end{align}
		From the RGE in Eq.~(\ref{eq:RGE-mass-matrix}) and Eq.~(\ref{eq:Q})
		one can identify
		\begin{align}
		a_{00}&=0, \nonumber\\
		a_{10}&=1+{\cal G}+\cdots,\nonumber\\
		a_{20}&=-\frac{1}{4}+\cdots,\nonumber\\
		a_{11}&=4+\cdots,
		\end{align}
		where the dots indicate contributions  from higher order $\beta$-functions to the corresponding structure $(P^T)^n M P^m$. 
		
		When the right-handed mass matrix is rank-2 at the cut-off scale, it suffices to consider terms up to ${\cal O}(P^2)$ in the Picard expansion. These depend on $a_{nm}$ with $n+m\leq 2$, although only $a_{11}$ is relevant for the calculation of the eigenvalues. When the right-handed mass matrix is rank-1 at the cut-off scale, the mass matrix at the scale $M_3$ is rank-2 when keeping terms up to ${\cal O}(P^2)$ and rank-3 when keeping terms up to ${\cal O}(P^4)$. Therefore, one may wonder whether it is necessary  a calculation of the three- and four-loop beta functions, which contribute to the terms ${\cal O}(P^3)$ and ${\cal O}(P^4)$ in the Picard expansion for a precise calculation of the neutrino mass spectrum. 
		
		To address this question we have calculated, using the full Picard expansion Eq.~(\ref{eq:Picard-approx}), the parameters $M_1$, $M_2$ and ${\mathbb P}'$ that determine the light neutrino mass matrix at the scale $\mu=M_3$ ({\it cf.} Eq.~(\ref{eq:mus})), including terms proportional to $a_{nm}$ with $n+m\leq 4$. We find that $M_2$ and ${\mathbb P}'$ do not change at the leading order in the expansion. On the other hand, $M_1$ reads:
		\begin{align}
		M_1\Big|_{\mu=M_3}&\simeq 8M_3\frac{\Big(P_{21}\left(P_{31}^2-P_{32}^2\right)-\left(P_{11}-P_{22}\right)P_{31}P_{32}\Big)^2}{P_{31}^2+P_{32}^2} \nonumber\\  &\times \left[\log^2\Big(\frac{M_3}{\Lambda}\Big) + \frac{4 a_{22} - a_{21}^2}{32}\log\Big(\frac{M_3}{\Lambda}\Big)\right],
		\end{align}	
		which amounts to a correction to Eq.~(\ref{eq:M1M2}) which we expect to be at most ${\cal O}(1)$ for $\log(\Lambda/M_3)\sim 1$, as assumed throughout the paper.

	\bibliographystyle{JHEP}
	\bibliography{references}

\end{document}